\documentclass[11pt]{article}       

\usepackage[a4paper,margin=2.5cm]{geometry}
\usepackage{graphicx}
\usepackage{dcolumn}
\usepackage{bm}
\usepackage{xcolor}
\usepackage[normalem]{ulem}
\usepackage{mathtools}
\usepackage{amsmath}
\usepackage{cancel}
\usepackage{bbold}
\usepackage{color}
\usepackage{dsfont}
\usepackage{braket}
\usepackage{latexsym}
\usepackage{bbm}
\usepackage{mathrsfs}
\usepackage{mathtools}
\usepackage{caption}
\usepackage{subcaption}
\usepackage{tikz}
\usepackage{tikz-feynman}
\usepackage{cite}

\newcommand{\CS}{\text{CS}}
\newcommand{\Tr}{\text{Tr}}

\usepackage[colorlinks]{hyperref}
\hypersetup{breaklinks=true}

\begin{document}

\title{\textbf{A new web of dualities from Majorana Fermions}}

\author{Andrea Amoretti\thanks{andrea.amoretti@ge.infn.it}\and Matteo Anselmi\thanks{matteoanselmi.phys@gmail.com} \and Daniel K. Brattan\thanks{danny.brattan@gmail.com}\\[5mm]
\small \emph{Dipartimento di Fisica, Università di Genova, via Dodecaneso 33, I-16146, Genova, Italy} \& \\
\small \emph{ I.N.F.N. - Sezione di Genova, via Dodecaneso 33, I-16146, Genova, Italy}}

\date{}

\renewenvironment{abstract}
 {\centerline{\bfseries Abstract}\vspace{4mm}}
 {}

\maketitle

\vspace{2cm}

\begin{abstract}
We explore new infrared dualities in $(2+1)$-dimensional quantum field theories involving Majorana fermions. 
Building on the recently proposed operator-deformation approach to bosonization dualities, we incorporate the bosonization of neutral fermions into the established web of Dirac dualities. 
Starting from the decomposition of a Dirac fermion into two Majorana components, we derive a hierarchy of dual bosonic descriptions based on orthogonal Chern--Simons--matter theories. 
This construction leads to novel boson--boson dualities, including an explicit correspondence between an $SO(N)$ theory with two real scalars and a $U(1)$ theory with one complex scalar, and their generalizations to arbitrary numbers of Majorana flavours. 
We analyze boundary conditions preserving chirality and demonstrate their role in matching gravitational anomalies and chiral central charges. 
Additional deformations yield orthogonal analogues of Ising--Gross--Neveu and pairing transitions. 
The resulting framework provides a unified Majorana-based extension of the three-dimensional duality web and clarifies the infrared structure of non-Abelian bosonization.

\end{abstract}

\newpage

\section{Introduction}

\noindent Dualities are a central theme in modern theoretical physics, providing complementary descriptions of the same physical phenomena and often uncovering deep structural relations between seemingly distinct theories. In quantum field theory, two models are said to be dual when their physical observables coincide, even though their microscopic degrees of freedom and Lagrangian formulations may differ drastically~\cite{Polchinski:2014mva}. Among the earliest and most influential examples is two-dimensional bosonization, first established by Coleman~\cite{Coleman:1974bu}, which identifies a massless Dirac fermion with a compactified scalar field through a precise mapping of operators and couplings.

In three dimensions, bosonization manifests as an \emph{infrared} (IR) duality, emerging at the long-distance fixed points of distinct microscopic theories. Two QFTs are IR-dual if their renormalization-group flows terminate at the same conformal field theory (CFT). Such dualities have attracted renewed interest~\cite{Karch:2016sxi, Seiberg:2016gmd, Witczak-Krempa:2023lgm, Metlitski:2016dht, Aharony:2016jvv}, owing both to their connection with strongly coupled CFTs and to their relevance in condensed-matter systems near quantum criticality \cite{Seiberg:2016gmd}\footnote{See also \cite{Amoretti:2013xya}, for the emergence of 3+1D Weyl fermions from bosonic salar tensor dualities.}.

A systematic approach to generating new IR dualities was proposed in~\cite{Witczak-Krempa:2023lgm}. Starting from the conjectured three-dimensional Dirac bosonization~\cite{Karch:2016sxi, Seiberg:2016gmd}, relevant operators are identified on both sides of the duality and used to deform the theories away from the critical point. When these deformations are promoted to dynamical scalar fields, the resulting renormalization-group flows lead to new interacting fixed points conjectured to be dual. The validity of these conjectures can be tested through the matching of global symmetries, low-energy degrees of freedom, and anomalies.

While the construction in~\cite{Witczak-Krempa:2023lgm} involves Dirac fermions, a parallel bosonization for \emph{Majorana fermions} has been proposed in~\cite{Metlitski:2016dht, Aharony:2016jvv}. It relates a single neutral Majorana spinor to an $SO(N)_1$ Chern--Simons theory coupled to an $N$-component real scalar field. At criticality, both theories describe the transition between a trivial insulator and a $p_x + i p_y$ topological superconductor, whose edge hosts a chiral Majorana mode~\cite{Stone:2003wbl, Gnezdilov:2015twa}.

The purpose of this work is to incorporate the Majorana bosonization into the existing duality web and to demonstrate that it naturally generates new classes of non-Abelian boson--boson equivalences. Starting from the two-flavour case, we derive a novel correspondence between an $SO(N)_1$ Chern--Simons theory coupled to two real scalars and a $U(1)_1$ theory coupled to a single complex scalar, with consistent boundary conditions ensuring chiral-central-charge matching. We then generalize the construction to an arbitrary even number of Majorana flavours, obtaining a hierarchy of dualities relating $SO(N)_1$ gauge theories to multiple copies of Abelian Chern--Simons--Higgs models. Finally, we study pairing-type deformations of the seed duality and show that they induce controlled Higgs flows $SO(N)\!\to\!SO(N{-}1)$, reproducing the expected thermal Hall responses in the IR.

\section{Generating new dualities from the Dirac seed}
An essential ingredient in our construction is the method developed in~\cite{Witczak-Krempa:2023lgm}. 
To illustrate its main features, we begin with the $(2+1)$-dimensional seed duality conjectured in~\cite{Karch:2016sxi, Seiberg:2016gmd}: 
\begin{subequations}
\label{Seed duality}
\begin{eqnarray} 
    \mathcal{L}_f[\psi;A]  \;&\longleftrightarrow & \;  \mathcal{L}_b[\phi,a;A] \; , 
\end{eqnarray}
where $\psi$ is a Dirac fermion, $\phi$ a complex scalar field, $A$ and $a$ two $U(1)$ gauge fields, and 
\begin{eqnarray}
    \label{Eq:FermionL}
    \mathcal{L}_f[\psi;A]  &=& i\Bar{\psi}\cancel{D}_A \psi \; , \\
    \label{Eq:BosonL}
    \mathcal{L}_b[\phi,a;A] &=&  |D_a\phi|^2 - |\phi|^4 + \frac{1}{4\pi} a \wedge \mathrm{d} a + \frac{1}{2\pi} A \wedge \mathrm{d} a \; . \qquad
\end{eqnarray}
\end{subequations}
 As noted in~\cite{Witczak-Krempa:2023lgm}, deforming the two sides by mass terms allows one to probe and match their distinct phases. 
On the bosonic side, a mass deformation 
	\begin{displaymath}
		\mathcal{L}_b[\phi,a;A] \rightarrow \mathcal{L}_b[\phi,a;A] - m_{b}^2 |\phi|^2	
	\end{displaymath}
 interpolates between a trivial gapped phase for negative mass and a Chern--Simons (CS) theory at level~$-1$ for positive mass ($m_{b}^2>0$). 
To reproduce the same low-energy behaviour the fermionic mass deformation, 
	\begin{displaymath}
		 \mathcal{L}_f[\psi;A]  \rightarrow  \mathcal{L}_f[\psi;A]  - m_{f} \Bar{\psi} \psi \; , 
	\end{displaymath}
 must have the opposite sign~\cite{Seiberg:2016gmd, Polyakov:1988md, Ferreiros:2018ohl},
\begin{align*}
    &m_f > 0 \quad \longleftrightarrow \quad m_b^{2} < 0 \, ,\\
    &m_f < 0 \quad \longleftrightarrow \quad m_b^{2} > 0 \, .
\end{align*}
Hence matching the mass deformation operators about \eqref{Seed duality} requires an overall minus sign,
\begin{equation}
    \Bar{\psi}\psi \;\longleftrightarrow\; -|\phi|^2 \, .
    \label{Mass terms identification Krempa seed duality}
\end{equation}
 
As usual for mass deformations, these operators are relevant and drive the two theories away from the critical point~\eqref{Seed duality}. 
Identifying pairs of relevant operators that correspond across the duality is the central idea of the method in~\cite{Witczak-Krempa:2023lgm}: by deforming both sides with the same operator, one moves along a shared RG trajectory and reaches new IR fixed points that can be conjectured to be dual as well.

More concretely, one introduces an auxiliary scalar field $\varphi$, coupled linearly to the relevant operators and endowed with a symmetry-breaking potential $v(\varphi)$. 
Applied to the seed duality~\eqref{Seed duality}, the deformation reads
\begin{equation}
    \mathcal{L}_f[\psi;A] - \varphi\,\Bar{\psi}\psi - v(\varphi)
    \;\longleftrightarrow\;
    \mathcal{L}_b[\phi,a;A] + \varphi\,|\phi|^2 - v(\varphi) \; ,
\end{equation}
where $\mathcal{L}_{f,b}$ are given in \eqref{Eq:FermionL} and \eqref{Eq:BosonL}, and the potential takes the usual Mexican-hat form,
\begin{equation}
    v(\varphi) = r\varphi^2 + \lambda\varphi^4 \, , \qquad \lambda>0 \, .
    \label{Scalar SSB potential}
\end{equation}
Promoting $\varphi$ to a dynamical field, with the proper kinetic term, means one can now explore the two regimes. 
For ${r>0}$, $\varphi$ is massive and can be integrated out, recovering the undeformed duality. 
For ${r<0}$, $\varphi$ develops a nonzero vacuum expectation value~$\varphi_0$, whose sign distinguishes the two massive phases discussed above. 
The critical point at ${r=0}$ therefore governs a transition between a trivial insulator and a topological phase described by a level-$(-1)$ CS theory.

\section{Bosonization of a Majorana fermion}
Our main goal is to incorporate Majorana fermions into this framework and to explore the resulting web of dualities. In the previous section we began with a massless theory and discussed how the mass deformation links the deformed theories; for the Majorana case let us begin with the full (conjectured) massive correspondence.
The bosonization of a single neutral fermion was first proposed in~\cite{Metlitski:2016dht, Aharony:2016jvv}, which relates the free theory of a Majorana spinor~$\xi$, 
\begin{equation}
    \mathcal{L}_\xi [\xi] = i\Bar{\xi} \cancel{\partial}\xi - m \Bar{\xi}\xi ,
    \label{Free Majorana action}
\end{equation}
 to a bosonic theory of an $N$-component real scalar field~$\vec{\phi}$ coupled to an $SO(N)$ Chern--Simons gauge field at level~one, 
\begin{eqnarray}
    S_b[\vec{\phi},a] &=& \int_M d^3x \, \mathcal{L}_{b}[\vec{\phi},a]  \nonumber \\
    		      &=& \int_M d^3x \left(|(\partial - i a)\vec{\phi}|^2 - r \vec{\phi} ^2 - \lambda (\vec{\phi}^2)^{2} \right) \\
    		      &\;& \hphantom{\int_M d^3x \left( \right.}+ \CS_{SO(N)}[a]_1 + N\CS_g \, . \nonumber
    \label{Bosonization of a free Majorana}
\end{eqnarray}
 This duality is expected to hold for ${N \ge 3}$. 
The Chern--Simons action, $\CS_{SO(N)}[a]_k$, is defined as
\begin{equation}
    \frac{k}{4\pi} \int_M d^3x \, \epsilon^{\mu\nu\rho}
    \!\left(a_\mu^a \partial_\nu a_\rho^a - \frac{2i}{3} f^{abc} a_\mu^a a_\nu^b a_\rho^c \right),
\end{equation}
while the gravitational contribution take the form
\begin{equation}
    \CS_g = \int_X \mathrm{Tr}\!\left(\frac{R\wedge R}{192\pi^2} \right),
    \label{Gravitational CS definition}
\end{equation}
where $X$ is a four-dimensional manifold with boundary ${\partial X = M}$, and $R$ denotes the Riemann curvature two-form.

Changing the sign of the Majorana fermion mass $m$ interpolates between two distinct phases: a trivial insulator, and a topological $p_x + ip_y$ superconductor with chiral central charge ${c_- = \tfrac{1}{2}}$. 
In the infrared, this correspondence takes the form
\begin{align}
    i\Bar{\xi}\cancel{\partial}\xi 
    \;\longleftrightarrow\;
    |(\partial - i a)\vec{\phi}|^2 - \lambda (\vec{\phi}^2)^2 + \CS_{SO(N)}[a]_1 + N\CS_g  ,
    \label{Majorana bosonization}
\end{align}
with the associated identification of relevant operators~\cite{Jian:2018amu}, 
\begin{equation}
    \Bar{\xi}\xi  \;\longleftrightarrow\;  -\vec{\phi}^2 \, .
    \label{Majorana mass term identification}
\end{equation} 

The construction extends naturally to $N_f$ flavours of Majorana fermions,
\begin{equation}
    S_\xi^{N_f} = \int_M d^3x \,\sum_{i=1}^{N_f}\big( i\Bar{\xi_i} \cancel{\partial}\xi_i - m \Bar{\xi_i}\xi_i \big) ,
\end{equation}
whose bosonic dual reads 
\begin{multline}
    S_{\phi}^{N_f}[\vec{\phi}_i,a] =
    \int_M d^3x \left[
    \sum_{i=1}^{N_f} |D_a \vec{\phi}_i|^2 
    - r\,\Tr M - u (\Tr M)^2 - v\,\Tr M^2 
    \right]\\
    + \CS_{SO(N)}[a]_1 + N\CS_g \, ,
    \label{Bosonization of multiple Majoranas}
\end{multline} 
where ${M_{ij} = \phi_i^\alpha \phi_j^\alpha}$, with $\alpha,\beta = 1,\dots,N$ labeling the fundamental representation of $SO(N)$ and $i,j = 1,\dots,N_f$ indexing the flavour space. 
The scalar potential takes the equivalent form
\begin{equation}
    V(\phi_i) = 
    r\, \phi_i^\alpha\phi_i^\alpha 
    + u\, (\phi_i^\alpha\phi_i^\alpha)(\phi_j^\beta\phi_j^\beta) 
    + v\, (\phi_i^\alpha\phi_j^\alpha)(\phi_j^\beta\phi_i^\beta) \, .
    \label{Scalar fields potential}
\end{equation}
Consistency with the single-flavour case subsequently requires
\begin{equation}
    N \ge N_f + 2 \, ,
    \label{N constraint}
\end{equation}
i.e. the number of colours must be greater than the number of flavours plus two.

\section{Dirac Fermion in terms of Majorana fermions}
Now towards the goal of our paper, a natural entry point for incorporating Majorana fermions into the known web of dualities is via their relation to Dirac particles. 
A Dirac spinor can be decomposed into two real components as
\begin{equation}
    \psi = \xi_1 + i\xi_2 \, ,
    \label{Dirac as two Majoranas}
\end{equation}
with each $\xi_i$ satisfying the Majorana condition~\cite{Tong:QFT}
\begin{equation}
    \xi_i = C\, \xi_i^* \, ,
\end{equation}
where $C$ denotes the charge-conjugation matrix in the two-dimensional spinor representation.  

The action of a free Dirac fermion on a three-manifold~$\mathcal{M}$ is
\begin{equation*}
    S = \int_\mathcal{M} d^3x \, \big(i\Bar{\psi}\cancel{\partial}\psi - m\Bar{\psi}\psi\big) \, .
\end{equation*}
Substituting~\eqref{Dirac as two Majoranas} gives
\begin{multline}
    \int_\mathcal{M} d^3x \, \Bar{\psi}(i\cancel{\partial}-m)\psi
    =  \int_\mathcal{M} d^3x \, \big[\Bar{\xi}_1(i\cancel{\partial}-m)\xi_1 + \Bar{\xi}_2(i\cancel{\partial}-m)\xi_2\big]
    + \int_{\partial\mathcal{M}} d^2x \,\Bar{\xi}_1 \gamma^\mu \xi_2 \, n_\mu \, ,
    \label{Dirac as two Majoranas action}
\end{multline}
where $n_\mu$ is the unit vector normal to~$\partial\mathcal{M}$. 
The boundary term arises from integrating by parts to obtain canonical kinetic terms for both Majorana components.  The corresponding equations of motion read
\begin{align}
    &\Bar{\xi}_1 : \quad i\gamma^\mu \partial_\mu\xi_1 + \gamma^\mu n_\mu \xi_2\, \delta(\partial \mathcal{M}) = 0 \, , 
    \label{Boundary EOM 1}\\[3pt]
    &\Bar{\xi}_2 : \quad i\gamma^\mu \partial_\mu \xi_2 = 0 \, ,
    \label{Boundary EOM 2}
\end{align}
where $\delta(\partial\mathcal{M})$ denotes localization at the boundary.

To elucidate the physical meaning of the boundary term, consider $\mathcal{M}$ as the half-space ${y < 0}$, with ${n_\mu = (0,0,1)}$ and ${\delta(\partial\mathcal{M}) = \delta(y)}$. 
Integrating~\eqref{Boundary EOM 1} from ${y=-\epsilon}$ to~$0$ and assuming that the integral of $i\gamma^0\partial_0\xi_1 + i\gamma^1\partial_1\xi_1$ vanishes as ${\epsilon\!\to\!0}$, one obtains
\begin{equation*}
    i\gamma^2 \xi_1\big|_{0} - i\gamma^2 \xi_1\big|_{0^-} + \gamma^2 \xi_2\big|_{0} = 0 \, .
\end{equation*}
Since $\gamma^2$ has no vanishing eigenvalues, this condition is equivalent to
\begin{equation}
    i\xi_1\big|_{0} + \xi_2\big|_{0} = i\,\xi_1\big|_{0^-} \, .
    \label{First boundary EOM}
\end{equation}
Applying the same procedure to~\eqref{Boundary EOM 2} yields the trivial continuity condition
\begin{equation*}
    \xi_2\big|_{0} = \xi_2\big|_{0^-} \, .
\end{equation*}

Boundary conditions can therefore be written as a linear combination of the two flavours vanishing at the edge:
\begin{equation}
    [a\,i\xi_1 + b\,\xi_2]\big|_{0} = 0 \, ,
    \label{Majorana boundary conditions}
\end{equation}
with ${a,b\in\mathbb{C}}$ nonzero. The choice ${b=-a}$ corresponds to the trivial condition ${\psi=0}$ on the boundary.

The boundary $\partial\mathcal{M}$ is $(1+1)$-dimensional, where chirality can be defined as in $(3+1)$ dimensions. 
Defining ${\gamma^5 = \gamma^0\gamma^1}$, which anticommutes with all $\gamma^\mu$, we can consider chiral boundary modes (eigenstates of $\gamma^5$) for the two Majorana flavours. 
These Majorana--Weyl fermions exist only in $(8k+2)$ spacetime dimensions~\cite{Polchinski:1998rr}. 
On the boundary we define
\begin{equation*}
    \gamma^5\xi_1^+ = \xi_1^+ \, , \qquad
    \gamma^5\xi_2^\pm = \pm \xi_2^\pm \, .
\end{equation*}
Applying $\gamma^5$ to~\eqref{Majorana boundary conditions} gives
\begin{gather*}
    \begin{cases}
        a\,i\xi_1^+ + b\,\xi_2^\pm = 0 \, ,\\[3pt]
        a\,i\xi_1^+ \pm b\,\xi_2^\pm = 0 \, ,
    \end{cases}
\end{gather*}
which admit non-trivial solutions only if the two Majorana components possess the same chirality. 
Consequently, the boundary conditions contribute a total chiral central charge
\[
    c_- = \pm 1 \, ,
\]
corresponding to two Majorana fermions of identical chirality.

This result is consistent with the Abelian Chern--Simons description of a Dirac fermion at low energies, which carries ${c_- = 1}$~\cite{Tong:2016kpv, Nawata:2022ywk}. 
As we will emphasize later, the non-trivial boundary structure will play a key role in matching the edge modes across dualities.

Finally, we note that these boundary conditions emerge naturally from the variational principle used to derive the equations of motion. 
Although boundary conditions are often imposed externally, this approach allows the theory itself to determine consistent constraints~\cite{Amoretti:2014iza,Amoretti:2014kba,Amoretti:2013nv}.

\section{Two Majorana flavours}
Given the results of the previous section, it is natural to begin with the bosonization of two flavours. 
At the critical point, the correspondence~\eqref{Majorana bosonization} implies an equivalence between
\begin{equation}
    \mathcal{L}_\xi[\xi_1,\xi_2]
    = i\Bar{\xi}_1\cancel{\partial}\xi_1 + i\Bar{\xi}_2\cancel{\partial}\xi_2 \, ,
\end{equation}
and 
\begin{equation}
    \mathcal{L}_\phi[\vec{\phi}_1,\vec{\phi}_2,a]
    = |D_a \vec{\phi}_1|^2 + |D_a \vec{\phi}_2|^2
    - \Tilde{V}(\vec{\phi}_{1},\vec{\phi}_{2})
    + \CS_{SO(N)}[a]_1 + N\CS_g \, ,
\end{equation}
 namely, 
\begin{equation}
    \label{eq:two_flavour_corresp}
    \mathcal{L}_\xi[\xi_1,\xi_2]
    \;\longleftrightarrow\;
    \mathcal{L}_\phi[\vec{\phi}_1,\vec{\phi}_2,a] \, .
\end{equation}
Here ${N \ge 4}$, and $\Tilde{V}(\vec{\phi}_{1},\vec{\phi}_{2})$ denotes the potential~\eqref{Scalar fields potential} without the quadratic term, which would move the theory away from the fixed point.

To analyze the phase structure, we follow the auxiliary-scalar method of~\cite{Witczak-Krempa:2023lgm}. 
Using the operator correspondence from~\cite{Metlitski:2016dht, Aharony:2016jvv, Jian:2018amu},
\begin{equation}
    \sum_{i=1}^{2} \Bar{\xi}_i\xi_i 
    \;\longleftrightarrow\;
    -\sum_{i=1}^{2} \vec{\phi}_i^2 \, ,
    \label{More flavours Majorana mass term identification}
\end{equation}
we deform both theories near criticality as 
\begin{align}
     \mathcal{L}_\xi[\xi_1,\xi_2]
    - \varphi \sum_{i=1}^2 \Bar{\xi}_i\xi_i - v(\varphi)   \;\longleftrightarrow\;
    \mathcal{L}_\phi[\vec{\phi}_1,\vec{\phi}_2,a]
    + \varphi \sum_{i=1}^2 \vec{\phi}_i^2 - v(\varphi) \, ,
\end{align} 
where $v(\varphi)$ takes the standard form~\eqref{Scalar SSB potential}.

The sign of~$r$ determines the two possible regimes. 
For ${r>0}$, $\varphi$ is heavy and can be integrated out, restoring the correspondence~\eqref{eq:two_flavour_corresp}. 
For ${r<0}$, $\varphi$ acquires a nonzero expectation value and induces mass terms for both fermions and scalars. 
On the bosonic side, the resulting quadratic term combines with $\Tilde{V}(\vec{\phi}_{1},\vec{\phi}_{2})$ to reproduce the full potential~$V(\vec{\phi}_{1},\vec{\phi}_{2})$ in~\eqref{Scalar fields potential}. 
Depending on the sign of ${\langle\varphi\rangle\equiv\varphi_0}$, the two theories describe either a trivial insulator or a topological phase with chiral central charge ${c_- = 1}$. 
These phases coincide with those found in~\cite{Witczak-Krempa:2023lgm}, as expected from the Dirac decomposition~\eqref{Dirac as two Majoranas action}. 

It is then natural to conjecture a duality between the corresponding bosonic theories: 
\begin{align}
     \sum_{i=1}^{2}|D_a \vec{\phi}_i|^2 - \Tilde{V}(\vec{\phi}_{1},\vec{\phi}_{2}) + \CS_{SO(N)}[a]_1 &+ N\CS_g \;\longleftrightarrow\; 
    |D_b z|^2 - |z|^4 + \CS[b]_1 \, ,
\end{align} 
where $\CS[b]_1$ denotes the Abelian CS term for the gauge field~$b$. 
The background gauge field coupled to the Dirac fermion in the seed duality has been turned off, since Majorana fermions are neutral. 
This correspondence can be summarized as
\begin{equation}
    SO(N)_1 + 2~\text{real scalars}
    \;\longleftrightarrow\;
    U(1)_1 + \text{one complex scalar} \, .
    \label{SO(N) and U(1)}
\end{equation}
Tracing the mass identifications gives
\begin{equation}
    \vec{\phi}_1^2 + \vec{\phi}_2^2 
    \;\longleftrightarrow\;
    |z|^2 \, .
    \label{Boson-boson duality mass identification}
\end{equation}

Applying the same deformation near the fixed point, 
 
\begin{equation}
    \mathcal{L}_\phi[\vec{\phi}_1,\vec{\phi}_2,a]
    - \varphi(\vec{\phi}_1^2+\vec{\phi}_2^2)
    - r \varphi^2 - \lambda \varphi^4
    \;\longleftrightarrow\;
    \mathcal{L}_z[z,b]
    - \varphi|z|^2 - r\varphi^2 - \lambda\varphi^4 \, ,
\end{equation}
  
we focus on the regime ${r<0}$. 
For ${\varphi_0>0}$, both $\vec{\phi}_i$ become massive, and integrating them out yields a pure $SO(N)$ CS theory with a gravitational term. 
On the right-hand side, $z$ also becomes massive, leaving a pure $U(1)_1$ gauge theory. 
While the bulk duality is consistent, the boundary requires care: the $U(1)$ theory carries ${c_-=-1}$, whereas the $SO(N)$ side is topologically trivial. 
The mismatch is resolved by the boundary term in~\eqref{Dirac as two Majoranas action}, whose consistent boundary conditions contribute precisely one unit of chiral central charge.  

The duality~\eqref{SO(N) and U(1)} should therefore be regarded as a correspondence between two pairs, each consisting of a bulk theory and a compatible set of boundary conditions. 
This is unsurprising, since the derivation originates from~\eqref{Dirac as two Majoranas}, where the two Majorana flavours are intrinsically linked.

For ${\varphi_0 < 0}$, the real scalars $\vec{\phi}_i$ condense, spontaneously breaking $SO(N)\!\to\!SO(N-2)$. 
On the $U(1)$ side, $z$ condenses and gives mass to $b$ via the Higgs mechanism, producing a trivial IR theory:
\begin{equation*}
    \CS_{SO(N-2)}[a]_1 + N\CS_g
    \quad \longleftrightarrow \quad 
    \text{trivial}\, .
\end{equation*}
Again, the edge theories differ by one unit of chiral central charge (${c_- = 1}$ on the left, ${c_- = 0}$ on the right), reconciled by the same boundary conditions preserving chirality. 
This transition connects two gapped phases with thermal Hall conductivities $0$ and $\pm \tfrac{\pi}{6}T$ at low temperatures~\cite{Kapustin:2019zcq, Zhang:2023hgq}.

It is illuminating to reinterpret these boundary conditions in the framework of~\eqref{SO(N) and U(1)}. 
Following~\cite{Metlitski:2016dht, Aharony:2016jvv}, the Majorana fermion can be identified with the monopole operator of the gauge field, consistent with viewing this bosonization as a relativistic form of flux attachment. 
For two flavours, this follows from the equation of motion for the temporal component of the gauge field $a^{i,0}$ derived from~\eqref{Bosonization of multiple Majoranas}:
\begin{equation*}
    \sum_{I=1}^{N_f=2}\rho_I^i - \frac{F^i}{2\pi} = 0 \, ,
\end{equation*}
where $\rho_I^i$ is the temporal component of the $SO(N)$ current for flavour~$I$, and $F^i$ is the non-Abelian field strength, with $i=1,\dots,\tfrac{N(N-1)}{2}$. 
Each scalar thus carries one unit of flux, as expected, and the operator mapping reads
\begin{equation}
    \xi_i
    \;\longleftrightarrow\;
    \vec{V}_M^i = \vec{\phi}_i\,\mathscr{M}_a \, ,
    \label{Majorana monopole identification}
\end{equation}
where $\mathscr{M}_a$ is the monopole operator of the gauge field~\cite{Turner:2019wnh}. 
The boundary condition~\eqref{Majorana boundary conditions} then translates to
\begin{equation*}
    [a \vec{V}_M^1 + b \vec{V}_M^2]\big|_{\partial\mathcal{M}} = 0 \, .
\end{equation*}
If monopoles are allowed on the boundary, ${\mathscr{M}_a|_{\partial\mathcal{M}}\neq 0}$, these reduce to scalar relations
\begin{equation*}
    [a \vec{\phi}_1 + b \vec{\phi}_2]\big|_{\partial\mathcal{M}} = 0 \, .
\end{equation*}
Conversely, the boundary conditions ensuring the consistency of~\eqref{SO(N) and U(1)} may be interpreted as forbidding monopoles at the edge,
\begin{equation*}
    \mathscr{M}_a\big|_{\partial\mathcal{M}} = 0 \, .
\end{equation*}

On the right-hand side of~\eqref{SO(N) and U(1)}, the analogous identification involves the Dirac fermion~\cite{Witczak-Krempa:2023lgm}:
\begin{equation*}
    \psi
    \;\longleftrightarrow\;
    z\,\mathscr{M}_b \, .
\end{equation*}
Hence, the boundary conditions for the Majorana operators can be viewed as relations between the real and imaginary components of this composite monopole operator. 
We therefore conjecture that, when supplemented by these boundary conditions, the duality~\eqref{SO(N) and U(1)} provides an alternative infrared description of the Chern--Simons--Higgs (CSH) model for ${N \ge 4}$.

\section{Even number of Majorana fermions}

The duality~\eqref{SO(N) and U(1)} extends naturally to an arbitrary even number of Majorana flavours. 
Consider the bosonization of $2N_f$ Majorana fields:
\begin{eqnarray}
    \mathcal{L}_\phi[\vec{\phi}_i,a]
    = \sum_{i=1}^{2N_f} |D_a \vec{\phi}_i|^2
      - \tilde V(\vec{\phi}_{1},\ldots,\vec{\phi}_{2N_{f}})\; + \CS_{SO(N)}[a]_1 + N\CS_g \, ,
\end{eqnarray}
with ${N \ge 2N_f+2}$. 
This can be related to the bosonic description of $N_f$ Dirac fermions,
\begin{equation}
    \mathcal{L}_z[z_i,b_i]
    = \sum_{i=1}^{N_f}\big(|D_{b_i}z_i|^2 - |z_i|^4 + \CS[b_i]_1\big) \, ,
    \label{Nf Dirac bosonization}
\end{equation}
and, in analogy with~\eqref{Boson-boson duality mass identification}, we identify the mass operators as
\begin{equation*}
    \sum_{i=1}^{2N_f} \vec{\phi}_i^2
    \;\longleftrightarrow\;
    \sum_{i=1}^{N_f} |z_i|^2 \, .
\end{equation*}

The deformation proceeds exactly as before. 
For ${\varphi_0>0}$, all scalars on the $SO(N)$ side are massive and can be integrated out, leaving a pure $SO(N)_1$ CS theory (plus the gravitational term). 
On the Abelian side, the $z_i$ are also massive, giving $N_f$ copies of a pure $U(1)_1$ CS theory. 
Once again, the bulk phases match while the edge central charges differ: ${c_-=0}$ for $SO(N)_1$ (after including the boundary contribution discussed previously) versus ${c_-=-N_f}$ for $N_f$ copies of $U(1)_1$. 
Appropriate boundary conditions preserving chirality supply the missing units and reconcile the edge data.

With multiple flavours there is additional freedom in choosing boundary conditions. 
One convenient option is a flavourwise pairing,
\begin{equation*}
    \big[a\,i\xi_{2n-1} + b\,\xi_{2n}\big]\Big|_{\partial\mathcal{M}} = 0 \, , \qquad n=1,\dots,N_f \, ,
\end{equation*}
but flavour-mixing choices are also possible. 
For instance, when ${N_f=2}$,
\begin{align*}
    &[a\,i\xi_1 + b\,i\xi_3]\Big|_{\partial\mathcal{M}} = 0 \, ,\\
    &[c\,\xi_2 + d\,\xi_4]\Big|_{\partial\mathcal{M}} = 0 \, ,
\end{align*}
and suitable linear combinations of the boundary equations of motion (analogous to~\eqref{First boundary EOM}) impose relations among different flavours, e.g.
\begin{equation*}
    a\,\xi_2\Big|_{\partial\mathcal{M}} + b\,\xi_4\Big|_{\partial\mathcal{M}}
    = a\,i\xi_1\Big|_{\partial\mathcal{M}^-} + b\,i\xi_3\Big|_{\partial\mathcal{M}^-} \, ,
\end{equation*}
with $\partial\mathcal{M}^-$ denoting a point infinitesimally inside the boundary.

In summary, the resulting IR duality is
 
\begin{equation}
    SO(N)_1 + 2N_f~\text{real scalars}
    \;\longleftrightarrow\;
    N_f\,\big(U(1)_1 + \text{one complex scalar}\big) \, ,
\end{equation}
 
conjectured to hold for ${N \ge 2N_f+2}$, with bulk phases matched by the deformation and edge central charges aligned by the same coherent boundary conditions used in the two-flavour case.

\section{Ising--Gross--Neveu}
\label{Sec: IGN}
Following Ref.~\cite{Witczak-Krempa:2023lgm}, a theory of two Dirac fermions with staggered mass terms exhibits a transition between a trivial insulator and a quantum spin Hall insulator~\cite{Cho:2010rk}. 
Once again, the use of Majorana variables provides a natural framework for constructing infrared dualities among purely bosonic theories.

In this setting we require four Majorana flavours. 
Since we will introduce two pairs of mass terms with opposite sign, it is convenient to apply twice the ${N_f=2}$ bosonization discussed previously, rather than a single ${N_f=4}$ construction.
The bosonic counterpart reads
 
\begin{multline}
    \mathcal{L}_\phi[\vec{\phi}_i,a,\tilde a]
    = \sum_{i=1}^{2} |D_a \vec{\phi}_i|^2 - \tilde V(\vec{\phi}_1,\vec{\phi}_2) + \CS_{SO(N)}[a]_1
       \\
      + \sum_{i=3}^{4} |D_{\tilde a} \vec{\phi}_i|^2 - \tilde V(\vec{\phi}_3,\vec{\phi}_4) + \CS_{SO(\tilde N)}[\tilde a]_1
      + (N+\tilde N)\CS_g \, .
      \label{IGN bosonization}
\end{multline}
 
Here, from~\eqref{N constraint}, ${N,\tilde N \ge 4}$, with no requirement that ${N=\tilde N}$. 
The notation makes explicit which fields enter each potential~$\tilde V$. 
On the Dirac side, the corresponding bosonization is given by Eq.~\eqref{Nf Dirac bosonization} with ${N_f=2}$. 
The relevant operators are identified using~\eqref{Boson-boson duality mass identification} as
\begin{equation*}
    \vec{\phi}_1^2 + \vec{\phi}_2^2 - \big(\vec{\phi}_3^2 + \vec{\phi}_4^2\big)
    \;\longleftrightarrow\;
    |z_1|^2 - |z_2|^2 \, .
\end{equation*}

The system enjoys a discrete $\mathbb{Z}_2$ symmetry exchanging the two sectors:
\begin{equation}
    \mathbb{Z}_2 :
    \begin{cases}
        1 \leftrightarrow 2 & \text{in } \mathcal{L}_z \, , \\[2pt]
        (1,2,a,N) \leftrightarrow (3,4,\tilde a,\tilde N) & \text{in } \mathcal{L}_\phi \, .
    \end{cases}
    \label{IGN Majorana Z2}
\end{equation}

We now perform the standard deformation with an auxiliary scalar $\varphi$ and analyze the resulting phases, which are interchanged by the $\mathbb{Z}_2$ symmetry~\eqref{IGN Majorana Z2}. 
For ${\varphi_0>0}$, on the Dirac (or $U(1)$) side, $z_2$ is massive while $z_1$ condenses, Higgsing $b_1$ and leaving the effective theory
\begin{equation*}
    \mathcal{L}_z^{\mathrm{eff}} = \CS[b_2]_1 \, .
\end{equation*}
On the $SO(N)\!\times\!SO(\tilde N)$ side, $\vec{\phi}_3,\vec{\phi}_4$ are massive, whereas $\vec{\phi}_1,\vec{\phi}_2$ condense and break $SO(N)\!\to\!SO(N-2)$, yielding
\begin{equation*}
    \mathcal{L}_\phi^{\mathrm{eff}} = \CS_{SO(N-2)}[a]_1 + \CS_{SO(\tilde N)}[\tilde a]_1 + (N+\tilde N)\CS_g \, .
\end{equation*}
The phase with ${\varphi_0<0}$ follows by applying the $\mathbb{Z}_2$ transformation. 
Once again, a mismatch in the edge central charges appears: the $U(1)$ side has ${c_-=-1}$, while the real-scalar side has ${c_-=1}$. 
As before, consistent boundary conditions involving four Majorana flavours contribute the missing two units of edge central charge, ensuring overall anomaly matching.

It is instructive to discuss the global symmetries. 
In addition to the discrete $\mathbb{Z}_2$ (spontaneously broken when $\varphi$ condenses), the Lagrangian $\mathcal{L}_\phi$ possesses an $SO(2)_\phi \times SO(2)_{\tilde\phi}$ flavour symmetry, acting as 
\begin{equation*}
    SO(2)_\phi :\quad \Phi_I \mapsto R_{IJ} \Phi_J \, , 
    \qquad \Phi=(\vec{\phi}_1,\vec{\phi}_2)^\top \, ,
\end{equation*} 
and analogously for $SO(2)_{\tilde\phi}$ acting on $\tilde\Phi=(\vec{\phi}_3,\vec{\phi}_4)^\top$. 
These are genuine global symmetries, distinct from the $U(1)^2$ gauge symmetry acting on the $z_i$. 
On the $U(1)$ side, the corresponding topological symmetries are $U(1)_T^2$, under which monopole operators carry charge~\cite{Turner:2019wnh}.

\section{Pairing transition}
\label{Sec: Pairing}
Another transition discussed in Ref.~\cite{Witczak-Krempa:2023lgm} involves a superconducting deformation of the Dirac seed duality~\cite{Karch:2016sxi}:
\begin{equation}
    \mathcal{L}_f
    = i\Bar{\psi}\cancel{D}_A \psi
      - \big(h\varphi\,\psi^2 + \text{c.c.}\big)
      - v(\varphi) \, ,
    \label{Seed deformed superconducting}
\end{equation}
where ${\psi^2 \equiv \epsilon^{\alpha\beta}\psi_\alpha\psi_\beta}$ with $\epsilon^{12}=-\epsilon^{21}=1$. 
On the bosonic side, using the identification of the Dirac operator with a monopole operator~\cite{Karch:2016sxi}, the quadratic operator maps as
\begin{equation}
    \psi^2 \;\longleftrightarrow\;
    \mathcal{M}_2 \equiv (z^\dagger)^2\,\mathscr{M}_a^{\,2} \, .
\end{equation}
The corresponding bosonic Lagrangian is 
\begin{eqnarray}
    \mathcal{L}_z
    = |D_a z|^2 - |z|^4
      + \frac{1}{4\pi} a \wedge da + \frac{1}{2\pi} A \wedge da \; - \big(h\varphi\,\mathcal{M}_2 + \text{c.c.}\big)
      - v(\varphi) \, .
\end{eqnarray} 

We now re-express this deformation in terms of Majorana degrees of freedom and then map it to the $SO(N)$ bosonic theory using the Majorana--monopole correspondence. 
Writing $\psi$ as in~\eqref{Dirac as two Majoranas} and adopting the conventions of App.~\ref{App: Majorana conventions}, one finds
\begin{equation}
    \psi^2 = -2\big[\chi_1\chi_1^* + i\chi_1\chi_2^* + i\chi_2\chi_1^* - \chi_2\chi_2^*\big] \, ,
\end{equation}
where $\xi_i = \begin{psmallmatrix}\chi_i\\ -\chi_i^*\end{psmallmatrix}$ denote the two Majorana flavours. 
Separating real and imaginary parts and defining the four-component field
\begin{equation}
    \xi \equiv \begin{pmatrix}\xi_1 \\ \xi_2\end{pmatrix} \, ,
\end{equation}
the bilinear takes the compact form
\begin{equation}
    \psi^2 = \Bar{\xi}\,S\,\xi \, , \qquad
    S =
    \begin{bmatrix}
        1 & 0 & i & 0 \\
        0 & 1 & 0 & i \\
        i & 0 & -1 & 0 \\
        0 & i & 0 & -1
    \end{bmatrix} .
\end{equation}
The fermionic theory becomes\footnote{As before, one must include the boundary term arising from integration by parts in the action~\eqref{Dirac as two Majoranas action}.}
\begin{equation}
    \mathcal{L}_f
    = i\Bar{\xi}\cancel{\partial}\xi
      - \big(h\varphi\,\Bar{\xi}S\xi + \text{c.c.}\big)
      - v(\varphi) \, .
\end{equation}
The background field $A$ is set to zero, as the Majoranas are neutral; the induced coupling
\begin{equation*}
    A_\mu
    \begin{pmatrix}\Bar{\xi}_1 & \Bar{\xi}_2\end{pmatrix}
    \begin{bmatrix}
        \gamma^\mu & i\gamma^\mu \\[2pt]
        -i\gamma^\mu & \gamma^\mu
    \end{bmatrix}
    \begin{pmatrix}\xi_1 \\ \xi_2\end{pmatrix}
\end{equation*}
admits no direct bosonic translation.

To connect with the $SO(N)$ theory, we use the Majorana--monopole mapping~\eqref{Majorana monopole identification} (for two flavours)
\begin{equation}
    \xi_i \;\longleftrightarrow\;
    \vec{V}_M^i \equiv \vec{\phi}_i\,\mathscr{M}_a \, .
\end{equation}
More generally,
\begin{equation}
    \xi_1 \;\longleftrightarrow\; (a \vec{\phi}_1 + b \vec{\phi}_2)\,\mathscr{M}_a \, , \qquad
    \xi_2 \;\longleftrightarrow\; (c \vec{\phi}_1 + d \vec{\phi}_2)\,\mathscr{M}_a \, ,
\end{equation}
with $a,b,c,d\in\mathbb{C}$ constrained by the mass-term identification~\eqref{More flavours Majorana mass term identification},
\begin{equation}
    |a|^2 + |c|^2 = |b|^2 + |d|^2 = 1 \, , \qquad
    a^*b + c^*d = 0 \, .
\end{equation}
Defining
\begin{equation}
    M \equiv \begin{pmatrix} \vec{\phi}_1\mathscr{M}_a \\ \vec{\phi}_2\mathscr{M}_a\end{pmatrix} ,
\end{equation}
the superconducting deformation maps to
\begin{equation}
    -\big(h\varphi\,M^\dagger S_N M + \text{c.c.}\big) \, , \qquad
    S_N =
    \begin{bmatrix}
        \mathbb{1}_{N\times N} & i\,\mathbb{1}_{N\times N} \\[2pt]
        i\,\mathbb{1}_{N\times N} & -\mathbb{1}_{N\times N}
    \end{bmatrix} .
\end{equation}
Since $\mathscr{M}_a$ is unitary~\cite{Turner:2019wnh}, 
\begin{equation}
    M^\dagger S_N M
    = \vec{\phi}_1^2 - \vec{\phi}_2^2
      + 2 i \vec{\phi}_1 \cdot \vec{\phi}_2  \, .
\end{equation} 
Writing $M^\dagger S_N M \equiv R + I$ and taking $h>0$,
\begin{equation}
    -\big(h\varphi\,M^\dagger S_N M + \text{c.c.}\big)
    = -2h\big(\Re\varphi\,R + i\,\Im\varphi\,I\big) \, .
\end{equation}
With $\varphi = \tfrac{1}{2}(\varphi_R + i\varphi_I)$, the full bosonic Lagrangian becomes 
\begin{multline}
    \mathcal{L}_\phi
    = |D_a \vec{\phi}_1|^2 + |D_a \vec{\phi}_2|^2 - \tilde V(\vec{\phi}_{1},\vec{\phi}_{2})
      - h\Big[\varphi_R\big(\vec{\phi}_1^2 - \vec{\phi}_2^2\big)
        - 2 \varphi_I \vec{\phi}_1 \cdot \vec{\phi}_2 \Big]\\
      - v(\varphi)
      + \CS_{SO(N)}[a]_1 + N\CS_g \, .
\end{multline}

For a complex auxiliary field,
\begin{equation}
    v(\varphi) = r|\varphi|^2 + \lambda|\varphi|^4 \, ,
\end{equation}
and for $r<0$ the field acquires a VEV $\varphi_0 = |\varphi_0|e^{i\theta}$, spontaneously breaking the $U(1)$ symmetry. 
While $|\varphi_0|$ sets an overall scale, the phase $\theta$ determines the relative weights of the two quadratic deformations.

For $\theta=0,\pi$ (real VEV), the IR Lagrangian becomes
\begin{equation}
    \mathcal{L}_\phi
    = |D_a \vec{\phi}_1|^2 + |D_a \vec{\phi}_2|^2
      - \tilde V(\vec{\phi}_{1},\vec{\phi}_{2})
      + \CS_{SO(N)}[a]_1+ N\CS_g
      - h\varphi_{R,0}\big(\vec{\phi}_1^2 - \vec{\phi}_2^2\big) \, .
\end{equation}
This breaks the $SO(2)_f$ flavour symmetry down to the $\mathbb{Z}_2$ acting as $(\vec{\phi}_1,\vec{\phi}_2)\mapsto(-\vec{\phi}_1,-\vec{\phi}_2)$. 
For $\theta=0$, $\vec{\phi}_1$ has positive mass $m_R^2 = h\varphi_{R,0}$ while $\vec{\phi}_2$ has negative mass of equal magnitude; the gauge symmetry is Higgsed to $SO(N-1)$, leaving the infrared theory
\begin{equation}
    \mathcal{L}_\phi\big|_{\theta=0}
    = \CS_{SO(N-1)}[a]_1 + N\CS_g \, .
    \label{Superconduting effective theory}
\end{equation}

For $\theta=\tfrac{\pi}{2},\tfrac{3\pi}{2}$ (purely imaginary VEV),
\begin{equation}
    \mathcal{L}_\phi
    = |D_a \vec{\phi}_1|^2 + |D_a \vec{\phi}_2|^2
      - \tilde V(\vec{\phi}_{1},\vec{\phi}_{2})
      + \CS_{SO(N)}[a]_1+ N\CS_g
      + 2 h\varphi_{I,0} \vec{\phi}_1 \cdot \vec{\phi}_2  \, .
\end{equation}
The quadratic part of the action reads
\begin{equation}
    S_\phi
    = \int d^3x \,\Big[
        \partial_\mu \vec{\phi}_1 \cdot \partial^\mu \vec{\phi}_1
        + \partial_\mu \vec{\phi}_2 \cdot \partial^\mu \vec{\phi}_2
        + 2 h\varphi_{I,0} \vec{\phi}_1 \cdot \vec{\phi}_2       + \dots\Big] \, .
\end{equation}
In $d=3$, $[\phi_i]=[\varphi]=\tfrac{1}{2}$, implying $[h]=\tfrac{3}{2}$ and defining $m_I^2 \equiv h\varphi_{I,0}$. 
Fourier transforming, the quadratic form becomes
\begin{eqnarray}
\label{eq:fouriertransform}
    S_\phi
    = \int \frac{d^3p}{(2\pi)^3}
      \begin{pmatrix}\vec{\phi}_1^{\mathrm{T}}(p) & \vec{\phi}_2^{\mathrm{T}}(p)\end{pmatrix}
      \begin{bmatrix}
        -p^2 & m_I^2 \\[2pt]
        m_I^2 & -p^2
      \end{bmatrix}
      \begin{pmatrix}\vec{\phi}_1(-p) \\ \vec{\phi}_2(-p)\end{pmatrix}
      \; + \dots \, .
\end{eqnarray}
with the argument $p$ indicating the Fourier transform of $\vec{\phi}_i$ and $\vec{\phi}_{i}^{\mathrm{T}}$ the vector transpose. From \eqref{eq:fouriertransform} the propagators can be easily computed:
\begin{equation}
    G(p)
    = -\frac{1}{(p^2-m_I^2)(p^2+m_I^2)}
      \begin{bmatrix}
        p^2 & m_I^2 \\[2pt]
        m_I^2 & p^2
      \end{bmatrix} .
\end{equation}
The propagator can be made diagonal switching to the basis 
\begin{equation}
    \begin{pmatrix} \vec{\phi}_- \\ \vec{\phi}_+\end{pmatrix}
    = \frac{1}{\sqrt{2}}
      \begin{pmatrix} \mathbbm{1}_{N} & -\mathbbm{1}_{N} \\ \mathbbm{1}_{N} & \mathbbm{1}_{N} \end{pmatrix}
      \begin{pmatrix} \vec{\phi}_1 \\ \vec{\phi}_2\end{pmatrix} ,
\end{equation} 
which yields
\begin{equation}
    G_D(p)
    = \begin{bmatrix}
        -\dfrac{1}{p^2+m_I^2} & 0 \\[2pt]
        0 & -\dfrac{1}{p^2-m_I^2}
      \end{bmatrix} ,
\end{equation}
revealing one massive and one tachyonic mode. 
As before, integrating out the heavy scalar and letting the tachyon condense Higgses $SO(N)\!\to\!SO(N-1)$, yielding the same infrared theory~\eqref{Superconduting effective theory}. 
For a general phase with both $\varphi_{R,0}$ and $\varphi_{I,0}$ nonzero, the eigenmasses are $\pm m^2 = \pm(m_R^2 + m_I^2)$.

\section{Conclusions}
In this work we have extended the operator-identification method of Ref.~\cite{Witczak-Krempa:2023lgm}, originally developed for Dirac fermions, to the framework of Majorana bosonization. Starting from the decomposition of a Dirac spinor into two neutral components, we analyzed how this structure translates within the $SO(N)$ Chern--Simons formulation of Ref.~\cite{Metlitski:2016dht, Aharony:2016jvv}. The correspondence between the two bosonizations naturally leads to an orthogonal (IR) description of the Chern--Simons--Higgs model, in which boundary conditions for Majorana flavours play a central role. In particular, chirality-preserving boundary conditions map to specific constraints on monopole operators: when scalar fields satisfy these relations, monopoles may exist at the boundary, whereas anomaly matching requires their absence otherwise.

Building on this framework, we identified new critical points and phase transitions. In Section~\ref{Sec: IGN} we described an orthogonal analogue of the transition between a trivial insulator and a quantum spin Hall phase, while in Section~\ref{Sec: Pairing} we analyzed a pairing-type deformation of the Dirac seed duality that produces a VEV-dependent Higgs phase. A distinctive feature of all these results is the \emph{flexibility in the rank of the gauge group} appearing in Majorana bosonization.

Our study also highlights the limitations of the operator-identification method: although it efficiently reveals new candidate fixed points, it does not by itself guarantee that the associated renormalization-group flows coincide. 
By considering a free theory of multiple Majorana flavours, indeed, problematic results regarding the flexibility in the rank of the gauge group and the number of bosonized fermions may appear. A more complete RG treatment could refine or constrain the parameter range and clarify possible mismatches between the bosonic and fermionic trajectories. Likewise, the consistency of reusing the $N_f=2$ bosonization twice in the Ising--Gross--Neveu construction remains an open technical question.

Future developments should focus on two main directions. First, the ambiguity in the rank of the gauge group within Majorana bosonization should be systematically investigated, possibly through explicit RG analyses or lattice realizations. Second, the framework could be expanded to include further critical points beyond the Dirac--Majorana connection. Exploring known topological phases supporting Majorana modes, such as Kitaev spin liquids~\cite{Kitaev:2005hzj}, or identifying new classes of IR dualities among orthogonal Chern--Simons--matter theories may provide deeper insight into the landscape of three-dimensional topological quantum field theories.

\appendix
\section{Majorana conventions}
\label{App: Majorana conventions}
We use the representation for gamma matrices:
\begin{equation}
    \gamma^0 = \sigma^3 , \hspace{0.5cm} \gamma^1 = i\sigma^1 , \hspace{0.5cm} \gamma^2 = i\sigma^2 \; .
\end{equation}
We can immediately verify that the charge-conjugation operator $C$ that satisfies:
\begin{equation}
    C^{-1} \gamma^\mu C = -(\gamma^\mu)^*
\end{equation}
is represented by $C=\gamma^2\gamma^0 = \begin{bmatrix}
    0 & -1 \\
    -1 & 0
\end{bmatrix}$. This gives us the structure of a Majorana spinor which obeys:
\begin{equation}
    \xi = C \xi^* \; .
\end{equation}
If $\xi = \begin{pmatrix}
    \chi_1 \\
    \chi_2
\end{pmatrix}$, where $\chi_i$ are Grassmann variables, then a Majorana fermion is:
\begin{equation}
    \xi = \begin{pmatrix}
        \chi \\
        -\chi^*
    \end{pmatrix} \; .
\end{equation}

\bibliography{Bibliography}

\end{document}